\documentclass[journal]{IEEEtran}
\usepackage{amsmath}
\usepackage{graphicx}
\usepackage{tikz}
\usepackage{color}
\usetikzlibrary{shapes,arrows}
\usepackage{subcaption}
\newcommand\blfootnote[1]{%
  \begingroup
  \renewcommand\thefootnote{}\footnote{#1}%
  \addtocounter{footnote}{-1}%
  \endgroup
}

\usepackage[style=numeric-comp ,backend=bibtex, maxbibnames=5]{biblatex}
\bibliography{library.bib}
\title{MOABB: Trustworthy algorithm benchmarking for BCIs}
\author{
  \IEEEauthorblockN
  {
    Vinay Jayaram\IEEEauthorrefmark{1}\IEEEauthorrefmark{2},
    Alexandre Barachant\IEEEauthorrefmark{3}
  }

  \IEEEauthorblockA
  {
    \IEEEauthorrefmark{1}
    Max Planck Institute for Intelligent Systems, 
    Department Empirical Inference, 
    T\"{u}bingen, Germany \\
    Email: vjayaram@tue.mpg.de}
  
  \IEEEauthorblockA
  {
    \IEEEauthorrefmark{2}
    IMPRS for Cognitive and Systems Neuroscience, 
    University of T\"{u}bingen, 
    T\"{u}bingen, Germany \\
  }
  \IEEEauthorblockA
  {
    \IEEEauthorrefmark{3}
    CTRL-Labs,
    New-York, USA \\
    Email: alexandre.barachant@gmail.com
  }
}
\begin{document}
\maketitle
\begin{abstract}
  BCI algorithm development has long been hampered by two major issues:
Small sample sets and a lack of reproducibility. We offer a solution
to both of these problems via a software suite that streamlines both
the issues of finding and preprocessing data in a reliable manner, as
well as that of using a consistent interface for machine learning
methods. By building on recent advances in software for signal
analysis implemented in the MNE toolkit, and the unified framework for
machine learning offered by the scikit-learn project, we offer a
system that can improve BCI algorithm development. This system is
fully open-source under the BSD licence and available at
https://github.com/NeuroTechX/moabb. To validate our efforts, we
analyze a set of state-of-the-art decoding algorithms across 12
open access datasets, with over 250 subjects. Our analysis confirms
that different datasets can result in very different results for
identical processing pipelines, highlighting the need for trustworthy
algorithm benchmarking in the field of BCIs, and further that many
previously validated methods do not hold up when applied across
different datasets, which has wide-reaching implications for practical
BCIs.


\end{abstract}
\section{Introduction}
Brain-computer interfaces (BCIs) have long presented the neuroscience methods
community with a unique challenge. Unlike in vision research, where one
has a database of images and labels, a BCI is defined by a signal
recorded from the brain and fed into a computer, which can be influenced in any
number of ways both by the subject and by the experimenter. As a result,
validating approaches has always been a difficult task. Number of channels,
requested task, physical setup, and many other features vary between the
numerous publically available datasets online, not to mention issues of
convenience such as file format and documentation. Because of this, the BCI
methods community has long done one of two things to validate an new approach:
Recorded a new dataset, or used one of few well-known, tried-and-true datasets.

Recording a new dataset, the ideal way to show that a proposed
method works in practice, presents problems for post-hoc analysis. Without
making data public, it is impossible to know whether offline
classification results are convincing or due to some coding issue or
recording artifact. Further, it is well-known that differences in
hardware \cite{Searle2000,Lopez-Gordo2014}, paradigm \cite{Allison2010}, and
subject \cite{Allison2010} can have large differences in the
outcome of a BCI task, making it very difficult to generalize findings
from any single dataset.

Over the years many datasets have been published online, and serve as an
attractive option when time or hardware do not permit recording a new one. In
the last year and a half, over a thousand journal and conference submissions
have been written on the BCI Competition III \cite{Blankertz2006,Schloegl2005}
and IV \cite{Tangermann2012} datasets. Considering that these datasets have been
available publically for over a decade, the true number of papers which validate
results against them is likely much higher. While it is impossible to deny
the impact these two datasets have had on the field, relying so heavily on
a small number of datasets -- with less than 50 subjects total -- exposes the field
to several important issues. In particular, overfitting to the setups offered
there is likely.

Lastly, and possibly most problematically, the scarcity of available code for
BCI algorithms old and new puts the onus on each individual lab to reproduce the
code for all other competing methods in order to make a claim to be comparable
with the 'state-of-the-art' (SOA). As a result, the vast majority of novel BCI
algorithm papers compare either against other work from the same lab, or old,
easily implementable standards such as CSP \cite{Koles1990} or
channel-level variances combined with a classifier of choice \cite{Garrett2003}.

Computer vision has solved this problem with enormous datasets like Imagenet
\cite{Deng2009} bundled with machine learning packages
(Tensorflow\cite{tensorflow2015-whitepaper}, PyTorch, and
Theano\cite{Paszke2017}). However, generating BCI data is often a very taxing
process both physically and mentally, and so it is not reasonable to create
datasets of such size. Rather, the field requires many different people
recording data in many contexts in order to create an appropriate benchmark. We
propose our platform, the MOABB (Mother Of All BCI Benchmarks) Project, as a
candidate for this application. The MOABB project consists of the aggregation of
many publicly available EEG datasets, converted to a common format and bundled
in the software package, as well as a collection of SOA algorithms. Using this
system researchers can to automatically benchmark those algorithms and run an
automated statistical analysis, making the process of validating new algorithms
painless and reproducible. The source code is written in Python and publically available
under the BSD licence at https://github.com/NeuroTechX/moabb.

As an initial validation of this project, we present results on the
constrained task of binary classification in two-class imagined motor
imagery, as that is the most widely used motor imagery paradigm and
allows us to demonstrate the process across the largest number of
datasets.  However, we note that this is only the first question we
attempt to answer in this field. The format allows for many other questions,
including different channel types (EEG, fNIRS, or other), multi-class
paradigms, and also transfer learning scenarios as described in
\cite{Jayaram2016}. 


\section{Methods}
Any BCI analysis is defined by three elements: A dataset, a context, and
a pipeline. Here we describe how all of these components
are dealt with within our framework, and how specifically we set the
options for the initial analyses presented here.

\subsection{Datasets}

Public BCI datasets exist for a wide range of user paradigms and
recording conditions, from continuous usage to single-session to
multiple-sessions-per-subject. Within the current MOABB project, we
have unified the access to many datasets, described in Table
\ref{tab:datasets}.

\begin{table*}[ht]
  \centering
  \begin{tabular}{l || c | c | c | c | c | c | c }
    Name & Imagery & \# Channels & \# Trials & \# Sessions & \# Subjects & Epoch & Citations \\ \hline
    Cho et al. 2017 & Right, left hand & 64 & 200 & 1 & 49 & 0-3s & \cite{Cho2017} \\
    Physionet & Right, left hand & 64 & 40-60 & 1 & 109 & 1-3s & \cite{Schalk2004, Goldberger2000} \\
    Shin et al. 2017 & Right, left hand & 25 & 60 & 3 & 29 & 0-10s & \cite{Blankertz2007, Shin2017} \\
    BNCI 2014-001 & Right, left hand & 22 & 144 & 2 & 9 & 2-6s & \cite{Tangermann2012} \\
    BNCI 2014-002 & Right hand, feet & 15 & 160 & 1 & 14 & 3-8s & \cite{Steyrl2016} \\
    BNCI 2014-004 & Right, left hand & 3 & 120-160 & 5 & 9 & 3-7.5s & \cite{Leeb2007} \\
    BNCI 2015-001 & Right hand, feet & 13 & 200 & 2/3 & 13 & 3-8s & \cite{Faller2012} \\
    BNCI 2015-004 & Right hand, feet & 30 & 70-80 & 2 & 10 & 3-10s & \cite{Scherer2015a} \\
    Alexandre Motor Imagery & Right hand, feet & 16 & 40 & 1 & 9 & 0-3s & \cite{Barachant2012a}\\
    Yi et al. 2014 & Right, left hand & 60 & 160 & 1 & 10 & 3-7s & \cite{Yi2014} \\
    Zhou et al. 2016 & Right, left hand & 14 & 100 & 3 & 4 & 1-6s & \cite{Zhou2016}\\
    Grosse-Wentrup et al. 2009 & Right, left hand & 128 & 300 & 1 & 10 & 3-10s & \cite{Grosse-Wentrup2009} \\
    \hline
    \centering\bf{Total}: & & & & & \bf{275}& &
    
\end{tabular}
    \caption{Dataset attributes}
    \label{tab:datasets}
\end{table*}

Adding new open-source datasets is also simple via the MNE toolkit
\cite{Gramfort2014,Gramfort2013}, which is used for all preprocessing and channel
selection. Any dataset that can be made compatible with their framework can
quickly be added to the set of data offered by this project. In addition, the
project offers test functions to ensure candidate code conforms to the software
interface.

\subsection{Context}

A \emph{context} is the set of characteristics that defines the
preprocessing and validation procedure. To go from a recorded EEG
time-series to a pipeline performance value for a given subject or
recording session, many parameters must be defined. First, trials need to be
cut out of the continuous signal and pre-processed, which is
possible in many different ways when taking into account parameters such as
trial overlap, trial length, imagery type, and more. Once the
continuous data is processed into trials, and these trials are fed
into a pipeline, the next question of how to create training and test sets,
and how to report performance, comes into play. We separate these two
notions in our software and call them the \emph{paradigm} and the
\emph{evaluation} respectively.

\subsubsection{Paradigm}

A paradigm defines how one goes from continuous data to trials for a standard
machine learning pipeline to deal with. While not an issue in image processing,
as each trial is just one image, it is crucial in EEG and biosignals processing
because most datasets do not have exactly the same events defined in the
continuous data. For example, many datasets with two-class motor imagery use
left versus right hand, while some use hands versus feet; there are also many
possible non-motor imageries. For any reasonable analysis the specific sort of
imagery or ERP must be controlled for, as they all have different
characteristics in the data and further are variably effective across subjects
\cite{Scherer2015a,Allison2010}. After choosing which events or imageries are
valid, the question comes to pre-processing of the continuous data, in the form
of ICA cleaning, bandpass filtering, and so on. These must also be identical for
valid comparisons across algorithm or datasets. Lastly, there are questions of
how to cut the data into trials: What is the trial length and overlap; or, in
the case of ERP paradigms, how long before and after the event marker do we use?
The answers to all these questions are summed up in the paradigm object.

\subsubsection{Evaluation}

Once the data is split into trials and a pipeline is fixed, there are many ways
to train and test this pipeline to minimize overfitting. For datasets with
multiple subjects recorded on multiple days, we may want to determine which
algorithm functions best in multi-day classification. Or, we may want to
determine which algorithm is best for small amounts of training data. It is easy
to see that there are many possibilities for splitting data into train and test
sets depending on the question to be answered, and these must be fixed
identically for a given analysis. Furthermore, there is the question of how to
report results. Multiclass problems cannot use metrics like the ROC-AUC which
provide unbiased estimates of classifier goodness in binary cases; depending on
things like the class balance, various other metrics have their own benefits and
pitfalls. Therefore this must also be fixed across all datasets, contingent on
the class of predictions the pipelines attempt to make. We define this as our
\emph{evaluation}.

\subsection{Pipeline}

We define a \emph{pipeline} as the processing that takes one from raw
trial-wise data into labels, taking both spatial filtering and
classification model fitting into account. A convenient API for
dealing with this kind of processing is defined by
scikit-learn \cite{Pedregosa2011}, which allows for easily definable
dimensionality reduction, feature generation, and model fitting. To
maximize reproducibility we allow pipelines to be defined either by
yaml files or through python files that generate the objects, but
force all machine learning models to follow the scikit-learn
interface. 

In essence, the MOABB combines the preceding components into a procedure that
takes a list of algorithms and datasets and trains each pipeline to each subject
or recording session independently in order to generate goodness-of-fit scores
such as accuracy or ROC-AUC. These scores can then be visualized and used for
statistical testing.

\section{Statistical Analysis}
\label{stats}

At the end of the MOABB procedure there are scores for every subject in every
dataset with every pipeline. The goal of this project is to synthesize these
numbers into an estimate of how likely it is that each pipeline out-performs the
other pipelines. However, even if imagery type and channel number were held
constant, differences in trial amount, sampling rate, and even location and
hardware mean that we cannot expect subjects across datasets to be naively
comparable. Therefore, we run independent statistical tests within each dataset
and combine the p-values afterwards. A secondary problem is that the difference
distribution for two algorithms within a given dataset is very unlikely to be
Gaussian. It is well-known that some subjects are BCI illiterate
\cite{Allison2010}, which implies that no pipeline can reliably out-predict
another one on that subset of subjects. Therefore, for large enough datasets,
the distribution of differences in pipeline scores is very likely to be at least
bimodal.

To deal with this issue while also keeping the framework running fast enough to
execute on a normal desktop, we use a mixture of permutation and non-parametric
tests. Within each dataset, either a one-tailed permutation-based paired t-test
(for datasets with less than 20 subjects) or a Wilcoxon signed-rank test is run for each pair of
pipelines, generating a p-value for the hypothesis that pipeline $a$ is bigger
than pipeline $b$ for each pair of pipelines. These p-values are combined via
Stouffer's method\cite{Stouffer1949}, with a weighting given by the square root of the number of
subjects as suggested in \cite{liptak1958combination}, to return a final p-value
for each hypothesis. Since each score is compared against $N_{pipelines}-1$
other scores for the same subject, we also apply Bonferroni correction to
protect against false positives. In order to determine effect size, we computed
the standardized mean difference within datasets and combined them using the
same weighting as was given to Stouffer's method.

\section{Experiment}

To show off the possibilities of this framework, we ran various well-known BCI
pipelines from across many papers in order to conduct the first big-data,
side-by-side analysis of the state of the art in motor imagery BCIs.

\subsection{Context}
For the paradigm, we choose to look at datasets including motor
imagery.  Motor imagery is the most-studied sort of imagery for BCIs
\cite{Yuan2014}, and we further limit ourselves to the binary case as
this has not yet been solved. For evaluations, we choose
within-session cross-validation, as this represents the best-case
scenario for any pipeline, with minimal non-stationarity. 

\subsubsection{Paradigm}

As there are many methods that show that multiple frequency bands can lead to
improved BCI performance\cite{KaiKengAng2008}, and further that discriminative
data is concentrated in the anatomical frequency bands, we test two
preprocessing pipelines: A single bandpass containing both the alpha and beta
ranges, from $8-35\text{Hz}$, and another from $8-35\text{Hz}$ in 4Hz
increments. All data was also subsampled to 128Hz, as the memory requirements
became prohibitive otherwise.

\subsubsection{Evaluation}
The evaluation was chosen to be within-session, as that minimizes the effect of
non-stationarity. As this is a binary classification task, the ROC-AUC score was
chosen as the metric to score 5-fold cross validation (the splits were kept
identical for all pipelines in a given subject). In comparison with the more
interpretable classification accuracy, the ROC-AUC is less sensitive to
imbalanced classes, which is important in this case where the datasets vary
heavily. In order to return a single score per subject, the scores from each
session were averaged when multiple sessions were present.

\subsection{Pipelines}

We implement a selection of pipelines from the BCI literature, as well as the
well-known standards of CSP + LDA and channel-level variances + SVM. Specific
implemented pipelines are in Table \ref{tab:pipelines}; all hyperparameters were
set via cross-validation.
\begin{table*}
  \centering
  \begin{tabular}{ l || p{6cm} | p{6cm} | c | }
    
    Name & Preprocessing & Classifier & Introduced in \\ \hline
    CSP + LDA & Trial covariances estimated via maximum-likelihood with unregularized common spatial patterns (CSP). Features were log variance of the filters belonging to the 6 most diverging eigenvalues & Linear Discriminant Analysis (LDA) & \cite{Koles1990} \\ \hline
    DLCSPauto + shLDA & Trial covariances estimated by OAS \cite{Chen2010} followed by
                     unregularized CSP. Features were log variance on the 6 top filters. & LDA with Ledoit-Wolf shrinkage of the covariance term  & \cite{Lotte2011} \\ \hline
    TRCSP + LDA & CSP with Tikhonov regularization, features were log variance on the 3 best filters for each class & LDA  & \cite{Lotte2011} \\ \hline
    FBCSP + optSVM & Filter bank of 6 bands between 8 and 35 Hz followed by OAS covariance estimation and unregularized CSP. Log variance from each of the 4 top filters from each sub-band were pooled and the top 10 features chosen by mutual information were used. & A linear support vector machine was trained with its regularization hyperparameter set by a cross-validated grid-search from $[0.01 100]$. & \cite{KaiKengAng2008} \\ \hline
    TS + optSVM & Trial covariances estimated via OAS then projected into the Riemannian tangent space to obtain features & Linear SVM with identical grid-search & \cite{Barachant2013} \\ \hline
    AM + optSVM & Log variance in each channel & Linear SVM with grid-search & N/A \\ \hline
    
  \end{tabular}
  \caption{Processing pipelines}
  \label{tab:pipelines}
\end{table*}


\section{Results}
\begin{figure*}
    \centering
    \includegraphics[width=\textwidth]{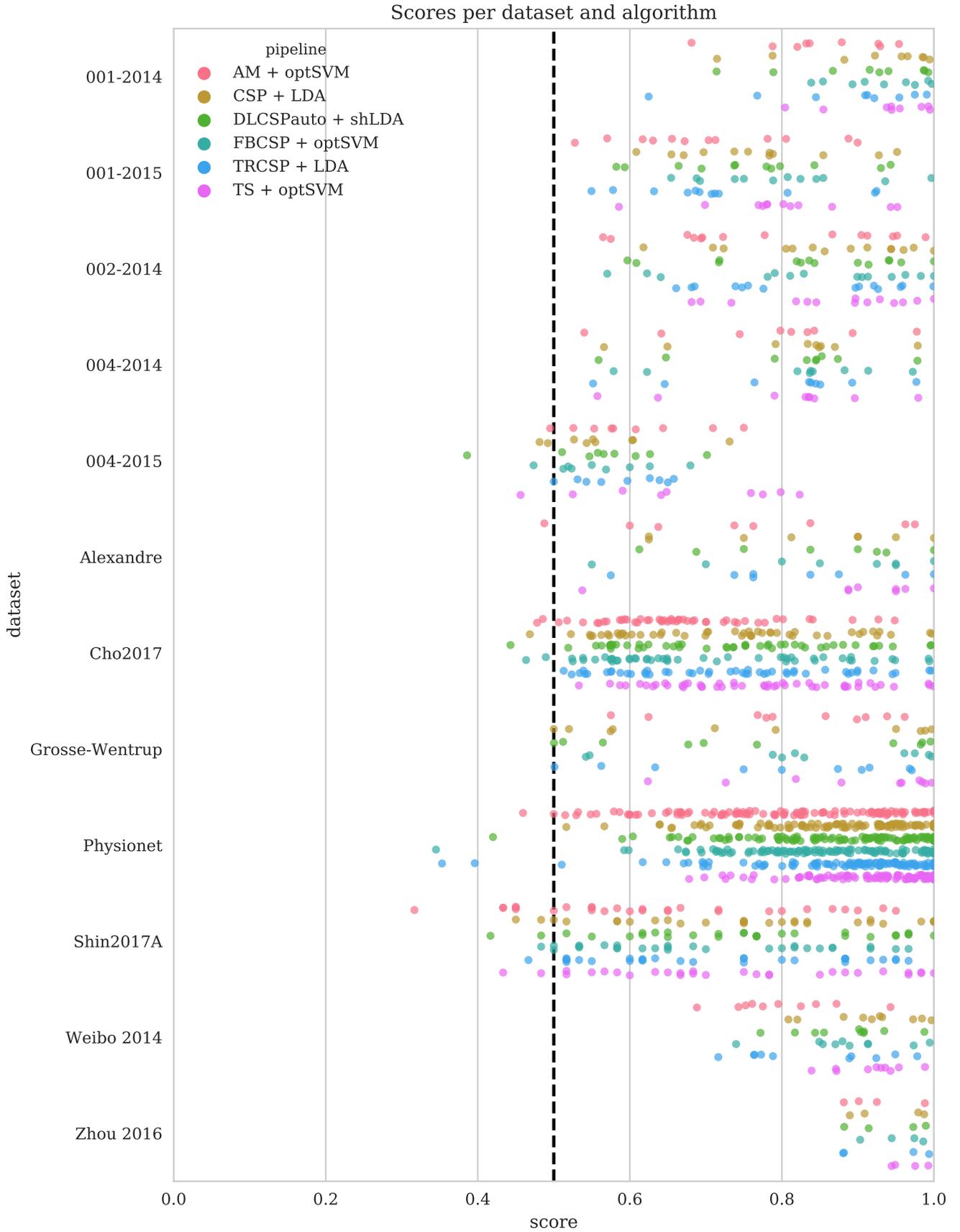}
    \caption{Visualization of all generated scores, across all datasets. The
      dotted line corresponds to a chance level performance of 0.5}
    \label{fig:all}
\end{figure*}

Figure~\ref{fig:all} shows all the results generated by this entire processing
chain. Surprisingly, perhaps, the pipelines do not clearly cluster on the
dataset level, making it unclear which ones perform best from simply this
plot. What is very clear, however, is that different datasets have very
different average scores independent of pipeline. This is particularly true when
one considers the case of \cite{Zhou2016} versus \cite{Goldberger2000}: Zhou et
al~\cite{Zhou2016} had pre-trained subjects, which compared to the naive sample
in the Physionet database makes a drastic difference.

\begin{figure*}
    \centering
    \begin{subfigure}[A]{0.5\textwidth}
        \includegraphics[width=\textwidth]{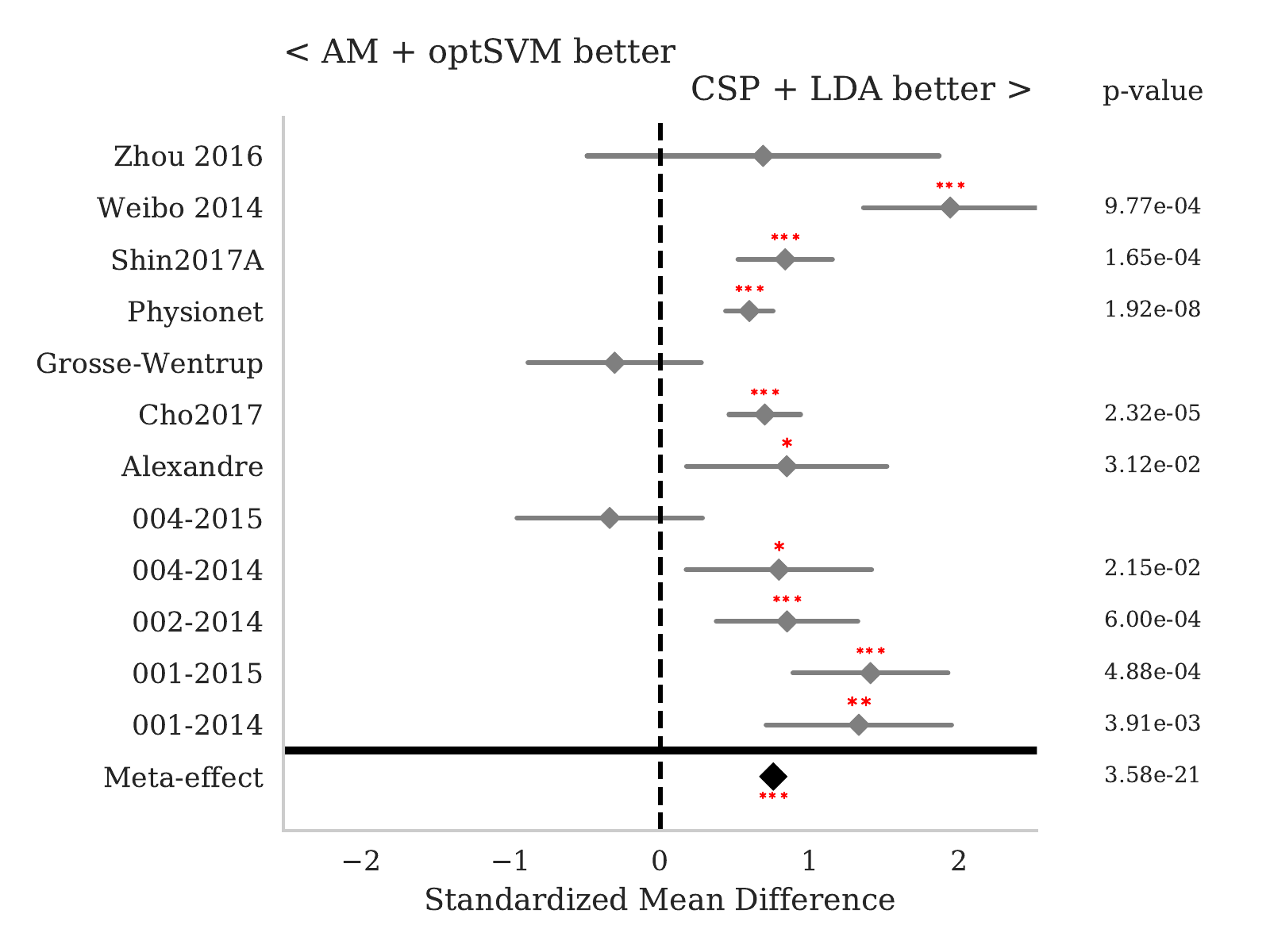}
    \end{subfigure}%
    \begin{subfigure}[B]{0.5\textwidth}
        \includegraphics[width=\textwidth]{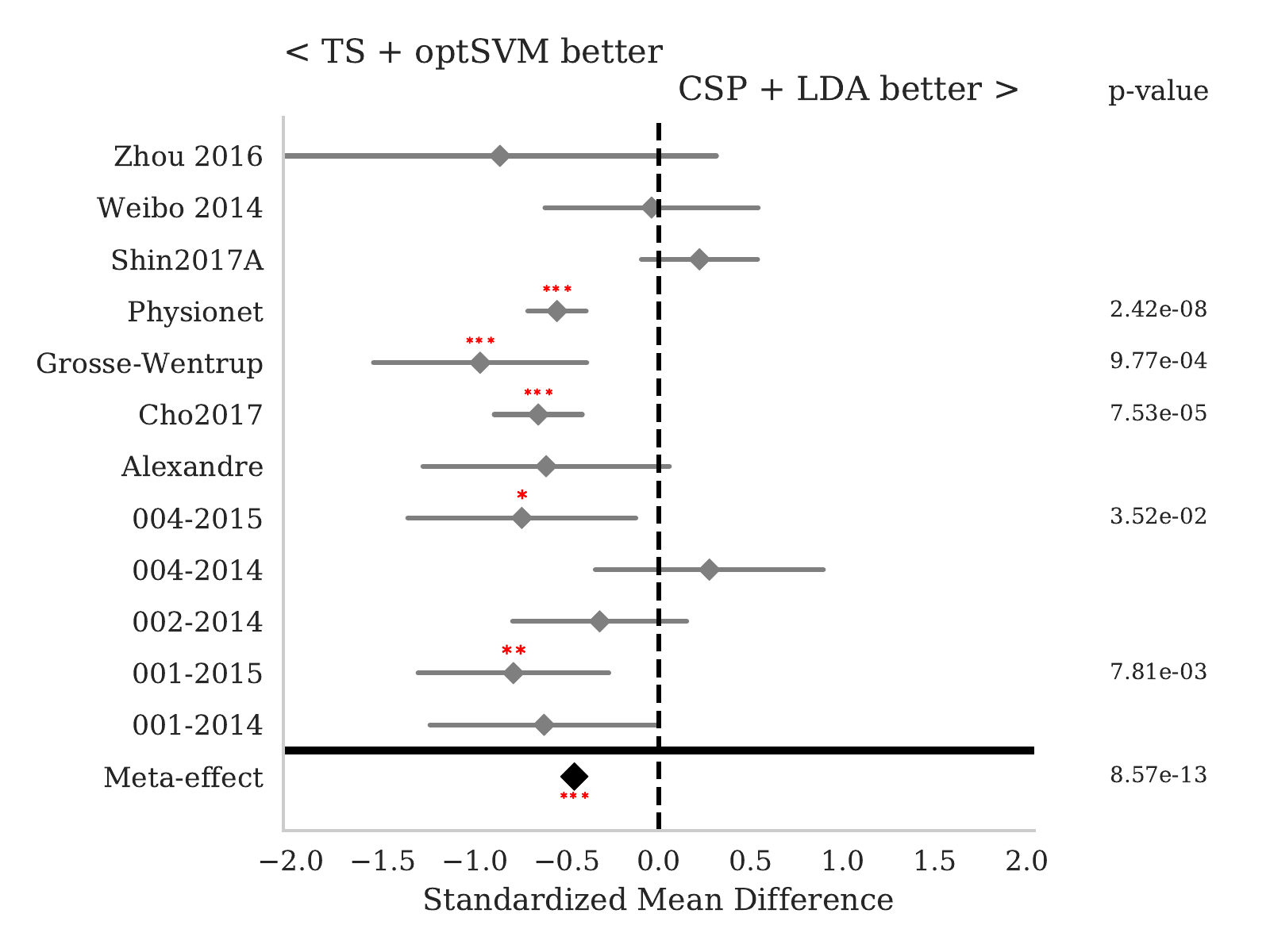}
    \end{subfigure}
    \caption{Meta-analysis style plots showing the performance of log variance
      features (A) and tangent space features (B) both compared against CSP. The
      effect sizes shown are standardized mean differences, with p-values
      corresponding to the one-tailed Wilcoxon signed-rank test for the
      hypothesis given at the top of the plot and 95\% interval denoted by the
      grey bar. Stars correspond to ***=p$<$0.001, **=p$<$0.01, *=p$<$0.05. The
      meta-effect is shown at the bottom of the plot. While there is a
      significant amount of variance between datasets--variance that could give
      contradictory results if these datasets were evaluated in isolation--the
      overall trend shows that CSP is on average better than channel
      log-variances and worse than tangent space projection.}
    \label{fig:vs_csp}
\end{figure*}

Figure~\ref{fig:vs_csp} shows the difference between CSP and the channel
log-variance and tangent space methods, as these are all well-known approaches
and have been compared against each other often in the past. Based on this
meta-analysis, CSP reliably out-performs channel log-variances across datasets
-- however, there are datasets such as \cite{Grosse-Wentrup2009} and
\cite{Scherer2015a} in which the opposite trend is shown. Similarly, while the
tangent space projection method normally out-performs CSP, that is also not true
for half of the sampled datasets. The confidence intervals also show why this is
likely the case -- for studies with very few subjects, such as \cite{Zhou2016},
the confidence intervals make even very strong standardized effects quite
untrustworthy.

\begin{figure*}
    \centering
    \begin{subfigure}[A]{0.33\textwidth}
        \includegraphics[width=\textwidth]{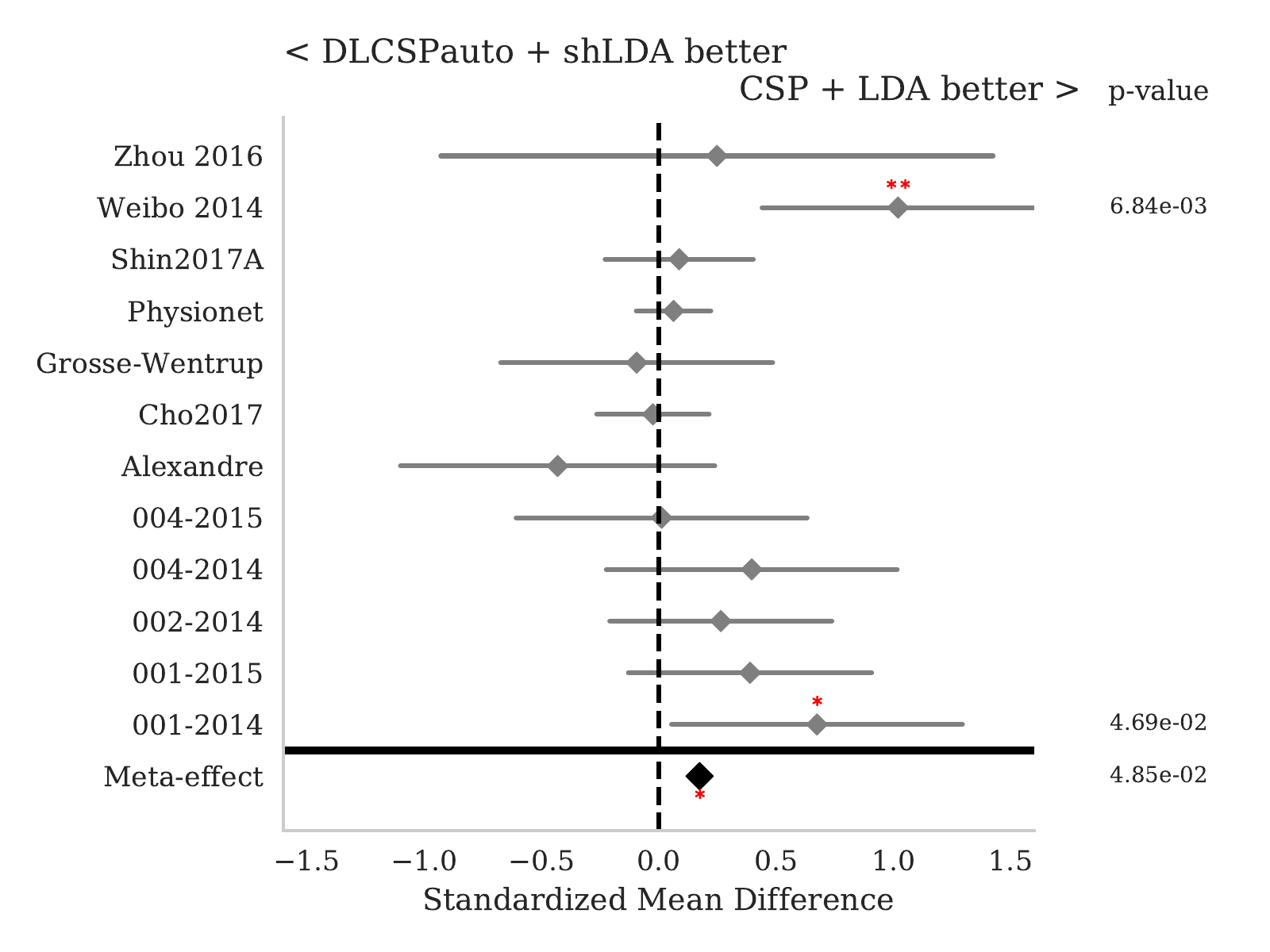}
    \end{subfigure}%
    \begin{subfigure}[B]{0.33\textwidth}
        \includegraphics[width=\textwidth]{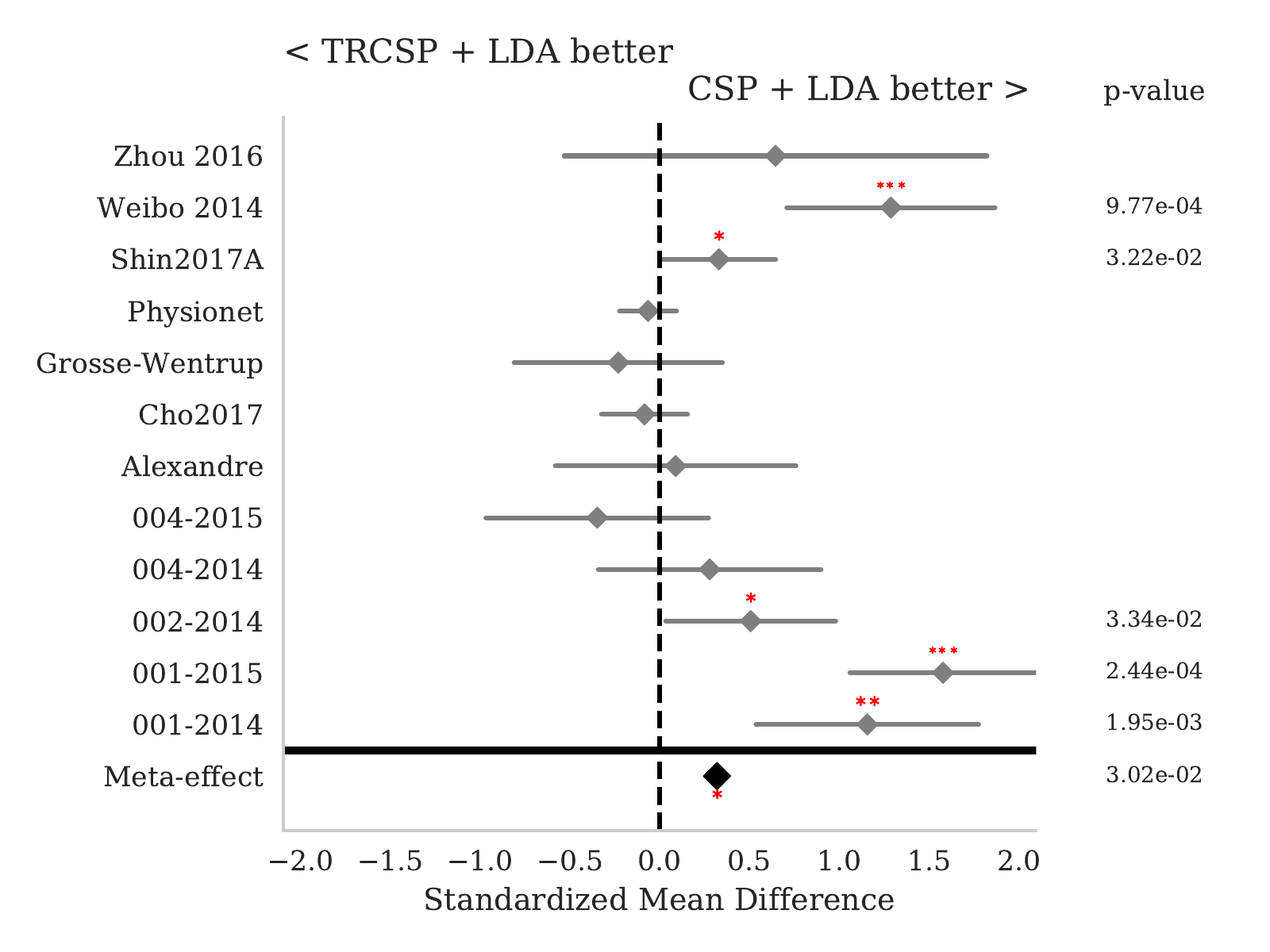}
    \end{subfigure}
    \begin{subfigure}[C]{0.33\textwidth}
        \includegraphics[width=\textwidth]{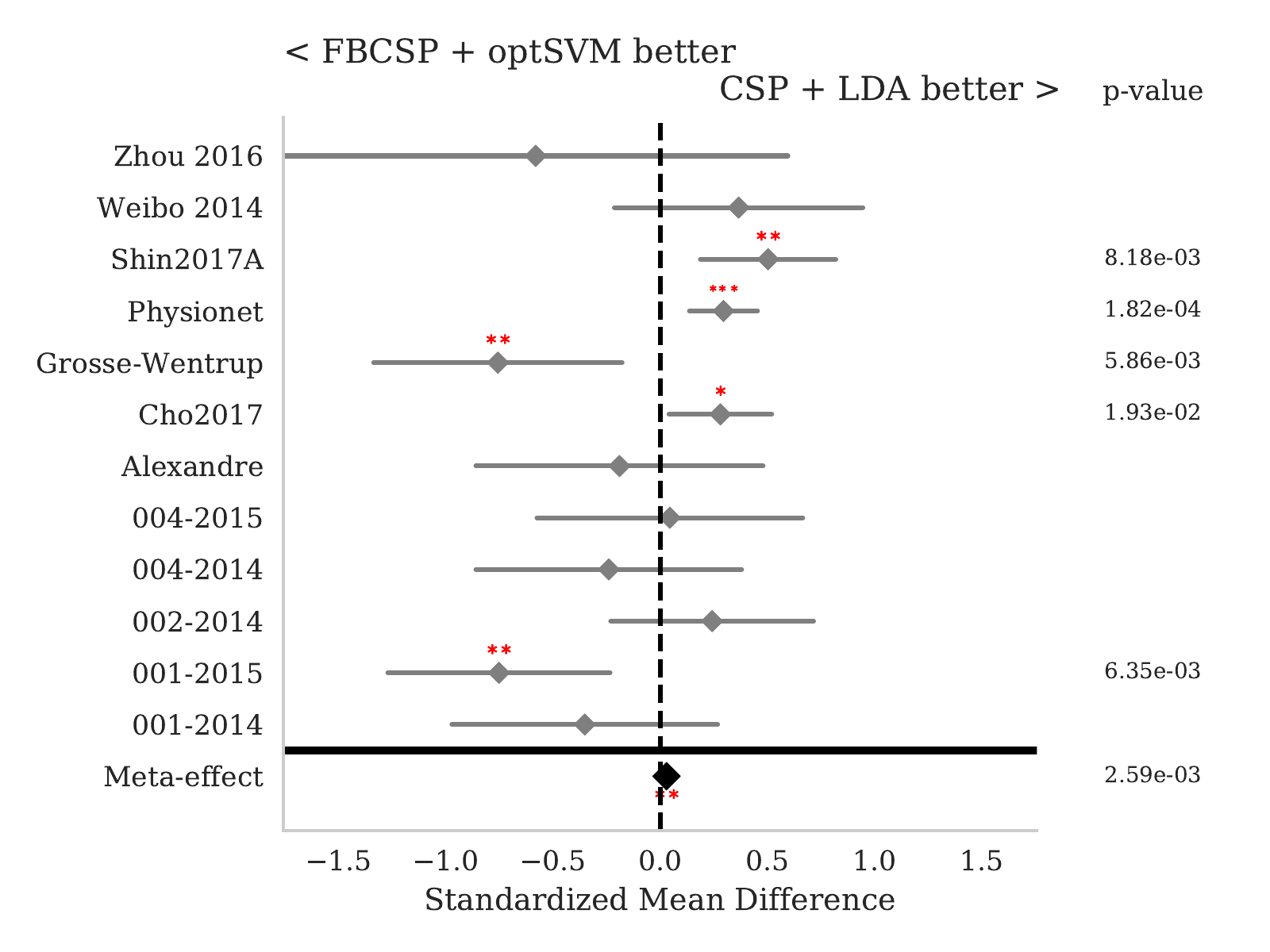}
    \end{subfigure}
    \caption{Meta-analysis style plots showing the performance of CSP versus CSP
      variants: DLCSPauto(A), TRCSP (B), and filter-bank CSP (C). The effect
      sizes shown are standardized mean differences, with p-values corresponding
      to the one-tailed Wilcoxon signed-rank test for the hypothesis given at
      the top of the plot and 95\% interval denoted by the grey bar. Stars
      correspond to ***=p$<$0.001, **=p$<$0.01, *=p$<$0.05. The meta-effect is shown
      at the bottom of the plot. While there is a significant amount of variance
      between datasets--variance that could give contradictory results if these
      datasets were evaluated in isolation--the overall trend shows that
      CSP out-performs the other algorithms in this setting.}
    \label{fig:csp_variants}
\end{figure*}

Figure~\ref{fig:csp_variants} compares CSP against commonly used variants. Here,
the difference is heavily dependent on dataset and no clear trend is visible. It
is interesting to note that in the case of filter-bank CSP, the BNCI 2014
datasets (which are included in the BCI Competition datasets used in
\cite{KaiKengAng2008}) show FBCSP to out-perform regular CSP while the opposite
is true for others such as Physionet. We further confirm the result from
\cite{Lotte2011} that regularizing the covariance estimates does not improve the
results of CSP. However, somewhat surprisingly, the finding that Tikhonov
weighting increases performance was not validated in this analysis.

\begin{figure*}
    \centering
    \includegraphics[width=\textwidth]{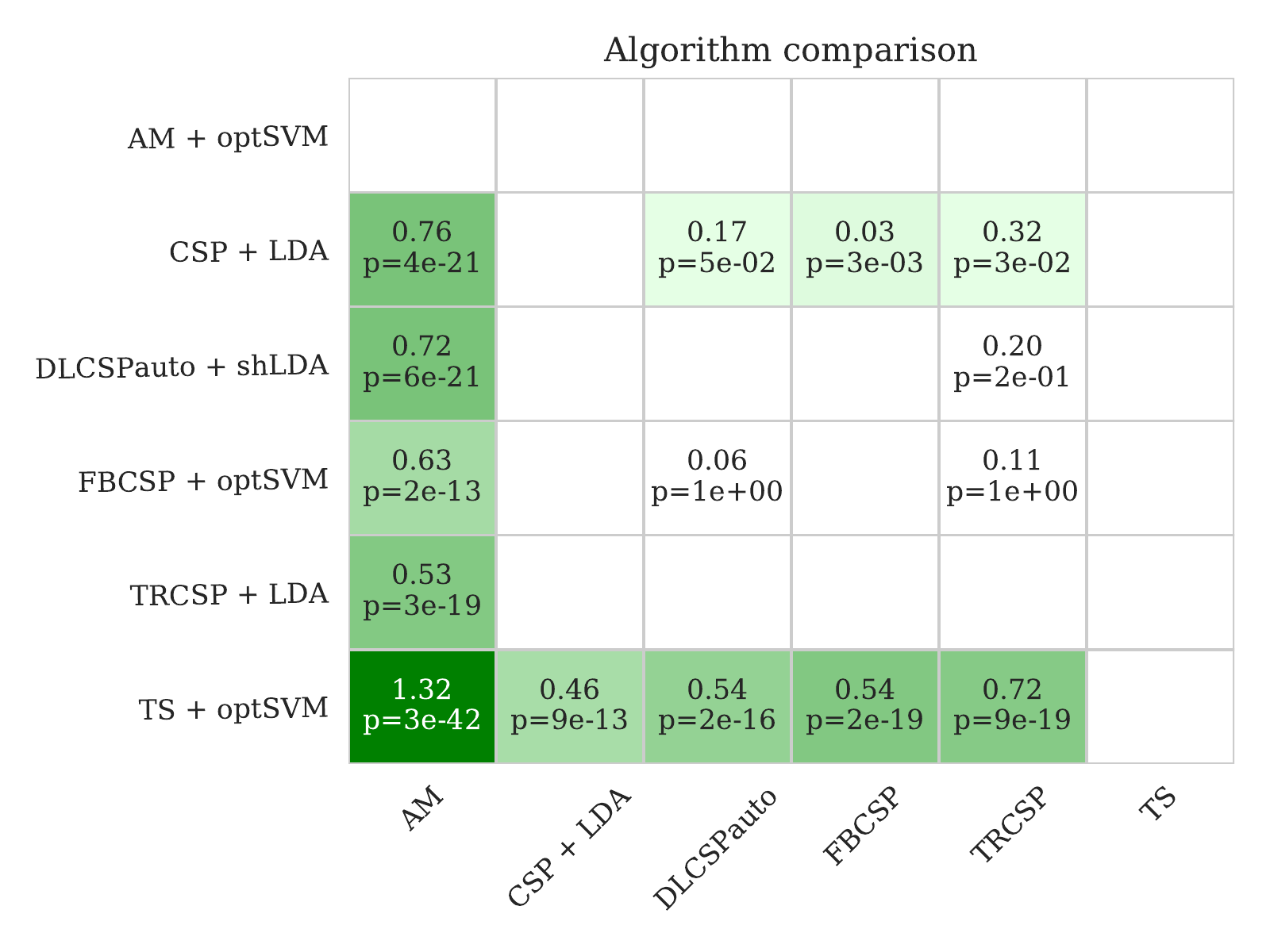}
    \caption{Ranking of algorithms in performance across all datasets with
      statistics generated as defined in section \ref{stats}. As all p-values
      are single-sided, in the case that the effect goes in the opposite
      direction of the hypothesis the values are removed for clarity. The
      values correspond to the standardized mean difference of the algorithm in
      the y-axis minus that in the x-axis and the associated p-value}
    \label{fig:rank}
\end{figure*}

The meta-effects shown in Figures~\ref{fig:vs_csp} and~\ref{fig:csp_variants}
are summed up in Figure~\ref{fig:rank}, which displays the meta-effect size in
cases that the algorithm on the y-axis significantly out-performed the algorithm
on the x-axis according to the statistical procedure outlined in Section
\ref{stats}, as well as the significance denoted by the stars under the
meta-effect size. Here we can see that all other algorithms out-performed
log-variance features on average (though with significant variance over datasets
as seen in the other figures) and that among CSP and its variants, tangent space
projection is better.

\section{Discussion}
We present a system for reliably comparing BCI pipelines that is both
easily extended to incorporate new datasets and equipped with an
automated statistical procedure for determining which pipelines
perform best. Furthermore, this system defines a simple interface for
submitting and validating new BCI pipelines, which could serve to
unify the many methods that exist so far. To test that system, we
present results using standard pipelines in contexts that have wide
relevance to the BCI community. By looking across multiple, large
datasets, it is possible to make statements about how BCIs perform on
average, without any sort of expert tuning of the processing chain,
and further to see where the major pitfalls still lie.

The results of this analysis suggest that many well-known methods do not
reliably out-perform simpler ones, despite the small-scale studies done years
ago to validate them. In particular, the world of CSP regularization literature
does not appear to have the effect that was originally claimed. Rather, the
major difference in BCI classification isn't actually the algorithm, as of now,
but the recording and human paradigm characteristics. The two most clear
findings to come out of this are that log variances on the channel level are
almost never better than CSP or Riemannian methods, and that the tangent space
classification pipeline is the best of the tested models for single-session classification.

In particular in the cases of FBCSP and the regularized approaches presented
here, the results presented here are surprising finding as they go against the
results reported in the original papers. In the case of FBCSP, we perform
similarly to the results shown in \cite{KaiKengAng2008}. BNCI 2014-001 and
2014-004 are originally from the BCI competitions and were used in the original
paper, and our finding is that on these datasets FBCSP indeed out-performed
regular CSP. In the case of the regularized variants DLCSPauto and TRCSP, our
results on the BCI competition data do not actually follow the originally
reported trend. Some possible for reasons are the following: our use of
single-session recordings ignores the initial training and test distinctions
given within the competition, and we also used the AUC-ROC instead of the
accuracy that was reported in the initial analysis. The full code to replicate
these results is available publically, and so we hope we can at least rule out
improper coding as a source of error.

Looking at these findings, it is particularly interesting to look at the case of
filter-bank CSP versus CSP, as in this analysis the significance goes in both
directions depending on the dataset. Since datasets vary in many
characteristics, such as channel number, imagery type, and trial time, it is
hard to determine what exactly underlies this diverging performance -- but it is
likely that this is not purely by chance. With increasing numbers of available
datasets, however, the answers to such differences become possible. If we have
many different situations in which to test algorithms, we can determine what
factors contribute to the differences in performance between them. It is also
important to emphasize that the results shown here must be taken in context. All
results were generated by cross-validation within single recording sessions,
which limits the possible non-stationarity. Because of this, regularization is
at its least useful -- which means that it would be inappropriate to dismiss
regularization in the case of CSP out of hand. Rather, this same analysis should
be re-run in the case of cross-session classification, a task that is currently
infeasible due to the number of multi-session datasets.

\section{Conclusion}

Meta-analysis is a well-described tool in other scientific fields to attempt to
synthesize the effects of many different studies that all bear on the same, or
very similar hypotheses. Though its use in BCIs has been hampered by the
difficulties involved in gathering the data and algorithms in a single place,
the MOABB project has the potential to offer a solution to this problem. The
analysis here, though done with over 250 subjects, is still only a fraction of
the number of subjects recorded for BCI publications over the years. With more
papers that describe more varied setups, the power of this system can only grow,
and what this analysis shows most clearly is that the sample size problem in
BCIs is bigger than we might have expected. By gathering the data and offering a
system for testing algorithms, we hope that this platform in the coming years
can help to solve it.


\section*{Acknowledgements}
We would like to extend our thanks to Dr. Marco Congedo for his valuable
input regarding the appropriate statistical procedure for this
analysis, and also to the NeuroTechX community for helping to get this
project started.

\blfootnote{corresponding author: Vinay Jayaram (email: vjayaram@tue.mpg.de)}
\printbibliography
\end{document}